\documentclass[sigconf]{acmart}

\AtBeginDocument{%
  }

\copyrightyear{2026}
\acmYear{2026}
\setcopyright{cc}
\setcctype{by}
\acmConference[MSR '26]{23rd International Conference on Mining Software Repositories}{April 13--14, 2026}{Rio de Janeiro, Brazil}
\acmBooktitle{23rd International Conference on Mining Software Repositories (MSR '26), April 13--14, 2026, Rio de Janeiro, Brazil}
\acmPrice{}
\acmDOI{10.1145/3793302.3793324}
\acmISBN{979-8-4007-2474-9/2026/04}

\usepackage{multirow}
\usepackage{makecell,pifont}

\usepackage{stfloats}

\begin{document}

\title{AnoMod: A Dataset for Anomaly Detection and Root Cause Analysis in Microservice Systems}

%
\author{Ke Ping}
\orcid{0009-0009-5433-4771}
\affiliation{%
  \department{Department of Computer Science}
  \institution{University of Helsinki}
  \city{Helsinki}
  \country{Finland}
}
\email{ke.ping@helsinki.fi}

\author{Hamza Bin Mazhar}
\orcid{0009-0009-6352-9810}
\affiliation{%
  \department{Department of Computer Science}
  \institution{University of Helsinki}
  \city{Helsinki}
  \country{Finland}
}
\email{hamza.mazhar@helsinki.fi}

\author{Yuqing Wang}
\orcid{0000-0003-0175-005X}
\affiliation{%
  \department{Department of Computer Science}
  \institution{University of Helsinki}
  \city{Helsinki}
  \country{Finland}
}
\email{yuqing.wang@helsinki.fi}

\author{Ying Song}
\orcid{0000-0001-9791-1879}
\affiliation{%
  \department{Department of Computer Science}
  \institution{University of Helsinki}
  \city{Helsinki}
  \country{Finland}
}
\email{ying.song@helsinki.fi}

\author{Mika V. Mäntylä}
\orcid{0000-0002-2841-5879}
\affiliation{%
  \department{Department of Computer Science}
  \institution{University of Helsinki}
  \city{Helsinki}
  \country{Finland}
}
\email{mika.mantyla@helsinki.fi}

%
\renewcommand{\shortauthors}{Ping et al.}

\begin{abstract}
Microservice systems (MSS) have become a predominant architectural style for cloud services. Yet the community still lacks high-quality, publicly available datasets for anomaly detection (AD) and root cause analysis (RCA) in MSS. Most benchmarks emphasize performance-related faults and provide only one or two monitoring modalities, limiting research on broader failure modes and cross-modal methods.
To address these gaps, we introduce a new multimodal anomaly dataset built on two open-source microservice systems: SocialNetwork and TrainTicket. We design and inject four categories of anomalies (Ano): performance-level, service-level, database-level, and code-level, to emulate realistic anomaly modes. For each scenario, we collect five modalities (Mod): logs, metrics, distributed traces, API responses, and code coverage reports, offering a richer, end-to-end view of system state and inter-service interactions. We name our dataset, reflecting its unique properties, as AnoMod.
This dataset enables (1) evaluation of cross-modal anomaly detection and fusion/ablation strategies, and (2) fine-grained RCA studies across service and code regions, supporting end-to-end troubleshooting pipelines that jointly consider detection and localization.
\end{abstract}

\ccsdesc[500]{Software and its engineering~Software reliability}
\ccsdesc[300]{Software and its engineering~Software performance}
\ccsdesc[300]{Software and its engineering~Cloud computing}

\keywords{Microservices, Anomaly Detection, Multimodal Dataset, AIOps}


\maketitle
\section{Introduction}
 
Nowadays, microservice architectures are popular in modern software development \cite{10.1145/3643991.3644890}.
However, it also poses a significant challenge for Site Reliability Engineers (SREs) to maintain system stability, as anomalies are inevitable given the system's complexity and distributed characteristics. 
Thereby, Artificial Intelligence for IT Operations (AIOps) emerged, aiming to automate anomaly detection and root cause analysis (RCA) to help SREs manage these systems effectively. 
Progress in AIOps research, however, is often hindered by the lack of high-quality, publicly available datasets that reflect real-world conditions \cite{10.1145/3715742, 10992557, 10.1145/3715005}.

Existing open-source multimodal datasets of MSS for anomaly detection and root cause analysis are summarized in Table~\ref{tab:datasets}. We found that these datasets have two main limitations. 
First, they cover only a limited range of anomaly types. Most datasets focus on performance-related anomalies such as resource contention and network latency, though a few also include other anomaly types such as code defects, API errors, and configuration issues. However, real-world MSS environments involve many more types of anomalies beyond these. For example, common anomalies occur at the database level (e.g., connection pool exhaustion \cite{10803340}), the service level (e.g., service response failure \cite{10.1145/3501297}), and the code level (e.g., logic errors or unhandled exceptions \cite{8580420}). 
Second, existing datasets \cite{10.1145/3611643.3616249, 10172617, 10.1145/3510003.3510180} simulate user requests but lack any measurement or reporting of service-level or API endpoint coverage. Consequently, it remains unclear whether the critical services or common endpoints are invoked during the simulated user requests, which may affect the representativeness of the resulting datasets.

Our work extends prior work and makes novel contributions in three aspects. First, we designed a taxonomy of anomaly cases across four levels to replicate common failure scenarios of MSS, following \cite{aws-well-architected-reliability-2024, azure-failure-mode-analysis, azure-service-fabric-testability, void-2024-report}. Second, guided by this taxonomy, we collect two additional data modalities: Code Coverage Report (CCR) and API Response (API Res). CCR shows which code statements and branches actually ran, 
while API Res reflect user-visible availability; together, they connect internal cause with external symptoms to support anomaly diagnosis. 
Third, when selecting anomaly injection targets, we prioritize services with the widest impact based on system dependency relationships, whereas prior works typically use random or predefined injection targets. In the mean time, by improving the API specification for EvoMaster, we achieve broader endpoint coverage and richer test interactions. The complete dataset is publicly available on Zenodo (DOI: \href{https://doi.org/10.5281/zenodo.18342898}{10.5281/zenodo.18342898}), and the collection scripts are available at \url{https://github.com/EvoTestOps/AnoMod}.

\begin{table*}[t]
\centering
\caption{Comparison of AnoMod dataset against Existing Monitoring Datasets for Microservice Systems}
\label{tab:datasets}
\begin{tabular}{@{}
  >{\raggedright\arraybackslash}p{1.65cm} 
  >{\raggedright\arraybackslash}p{3.0cm} 
  >{\raggedright\arraybackslash}p{3.0cm} 
  >{\raggedright\arraybackslash}p{4.0cm}  
  >{\raggedright\arraybackslash}p{4.0cm}  
@{}}
\toprule
\textbf{Dataset} & \textbf{MSS} & \textbf{Data Modality} & \textbf{Anomaly Cases} & \textbf{Workload and Anomaly Injection Setup} \\
\midrule
Nezha \cite{10.1145/3611643.3616249}
& TrainTicket, OnlineBoutique
& Logs, Metrics, Traces
& Performance (2); Code Defects (1)
& User operations \& Inject into random service \\
\addlinespace
DeepTraLog \cite{10.1145/3510003.3510180}
& TrainTicket
& Logs, Traces
& Interaction (6); Configuration (4); Code defects (4)
& Built-in testing scripts \& Faulty versions of TrainTicket \\
\addlinespace
Eadro \cite{10172617}
& TrainTicket, SocialNetwork
& Logs, Metrics, Traces
& Performance (3)
& Built-in testing scripts \& Inject into random service \\
\addlinespace
AIOps'21 \cite{li2022constructing}
& Commercial Bank Systems
& Logs, Metrics, Traces
& Performance (7)
& Real production workloads \& Not specified \\
\addlinespace
RCAEval \cite{10.1145/3701716.3715290}
& TrainTicket, OnlineBoutique, SockShop
& Logs, Metrics, Traces
& Performance (6), Code Defects (5)
& Random load across all services \& Inject into random service \\
\addlinespace
Multi-source \cite{10.1007/978-3-030-44769-4_13}
& OpenStack
& Logs, Metrics, Traces
& Service (3)
& Rally \& Injected into core control-plane components \\
\addlinespace
LO2 \cite{10.1145/3727582.3728682}
& Light-OAuth2
& Logs, Metrics
& API errors
& Locust \& Not specified \\
\addlinespace
\hline
AnoMod
& TrainTicket, SocialNetwork
& Logs, Metrics, Traces, CCR, API Res
& Performance (3); Service (3); Database (3); Code Logic (3)
& EvoMaster \& Injected into critical and high-impact services \\
\bottomrule
\end{tabular}%
\end{table*}

\section{Dataset Construction}

\subsection{Target System}
We build our dataset on two widely used open-source MSS: SocialNetwork \cite{10.1145/3297858.3304013} and TrainTicket \cite{8580420}.
Both systems have well-documented architectures and service interactions, allowing us to map each anomaly case to specific system components and thus inject anomalies systematically. SocialNetwork is a social media application that consists of 21 services. 
TrainTicket is a larger and more complex railway ticketing system with 41 microservices. 

\subsection{Anomaly Design}
To ensure our dataset reflects real-world failure scenarios, our anomaly design is grounded in industry practice \cite{void-2024-report, aws-well-architected-reliability-2024, azure-failure-mode-analysis, azure-service-fabric-testability}. 
According to large scale incident studies and cloud reliability guidance, production failures are not only related to resource exhaustion but also to service dependency interactions, data store behavior, network partitions, and application logic errors \cite{void-2024-report, 10803340, 10.1145/3501297, 8580420}. Focusing on performance injection would miss these important failure modes. 

Motivated by these insights, we design anomaly cases organized into four levels that systematically cover the system stack: For instance, Database-level cases such as Connection Pool Exhaustion represent capacity- and constraint-driven resource exhaustion scenarios (e.g., limits on concurrent database connections) discussed in reliability best practices \cite{aws-well-architected-reliability-2024}. Meanwhile, Service-level cases such as Response Failures emulate dependency-related degradations where client-side timeouts are necessary to prevent prolonged waiting and cascading impact in synchronous interactions \cite{aws-well-architected-reliability-2024, azure-failure-mode-analysis}. We further follow cloud testing guidance that advocates inducing real-world faults during workloads to validate recovery paths and system behavior under both graceful and ungraceful faults \cite{azure-service-fabric-testability}. We organized these validated scenarios into four levels:
(1) Performance level anomalies capture resource related issues; 
(2) Service level anomalies arise during inter-service communication or request handling; 
(3) Database level anomalies include data layer disruptions, reflecting the dependency failures often observed in production environments where data stores become system bottlenecks;
(4) Code level anomalies target logic or implementation related defects within microservices, thereby emulating software defects that are difficult to expose through infrastructure-level testing.

\subsection{Collected Multimodal Data}

Our dataset captures multimodal data and provides a multi-view of system behavior to give a deep understanding of normal and anomaly.
We structure these modalities into categories: monitoring data common in modern microservice systems, and behavior data offering deeper insights into code execution and external interactions.

\subsubsection{Monitoring Data}
\textbf{Logs} are collected from application containers, providing timestamped records of execution flows, errors, and context for debugging and postmortem analysis.
\textbf{Metrics} are time-series data collected via Prometheus \cite{prometheus}, covering node-level (resources), application-level (request counts), and process-level indicators. They aggregate system health and performance trends to support anomaly detection.
\textbf{Traces}, captured using Jaeger \cite{jaeger} and SkyWalking \cite{skywalking}, track request propagation across microservices. They expose dependencies, latency, and call-chain structure, essential for diagnosing distributed performance issues.

\subsubsection{Behavior Data}
To enrich the dataset beyond standard monitoring, we introduce two additional modalities that bridge runtime behavior with code execution and user visible outcomes.
\textbf{Code coverage reports} generated using language appropriate instrumentation (Gcov \cite{gnu-gcov-2024} for C++; JaCoCo \cite{jacoco} for Java). This modality reflects runtime activity directly to source code lines, branches, or functions. Its primary value lies in identifying which specific code path, including error handlers and rarely executed logic, is activated during normal operation versus anomaly cases. It may be of high value in testing environments and debugging as it provides direct evidence for code level root cause analysis process. 
\textbf{API Responses} capture the direct outcomes of client-service interactions, providing a crucial black-box perspective on system behavior. Collected through client side observation during workload execution, these records include essential details, namely, HTTP status codes, end-to-end latency, and response context (headers and body content). API Res directly reflect user visible symptoms. They serve as an important link between low level system telemetry and the overall service reliability experienced by users, offering high level indicators for anomaly detection. 

Data from these five modalities, collected synchronously during controlled experiments, provide a rich, multiple perspective dataset. The monitoring data offers runtime visibility, while the behavior data supplements crucial context about internal code execution and external service outcomes, enabling more comprehensive multimodal correlation for anomaly detection and root cause analysis.

\subsection{Data Collection Pipeline}
\begin{figure}[t]
  \centering
  \includegraphics[width=\linewidth]{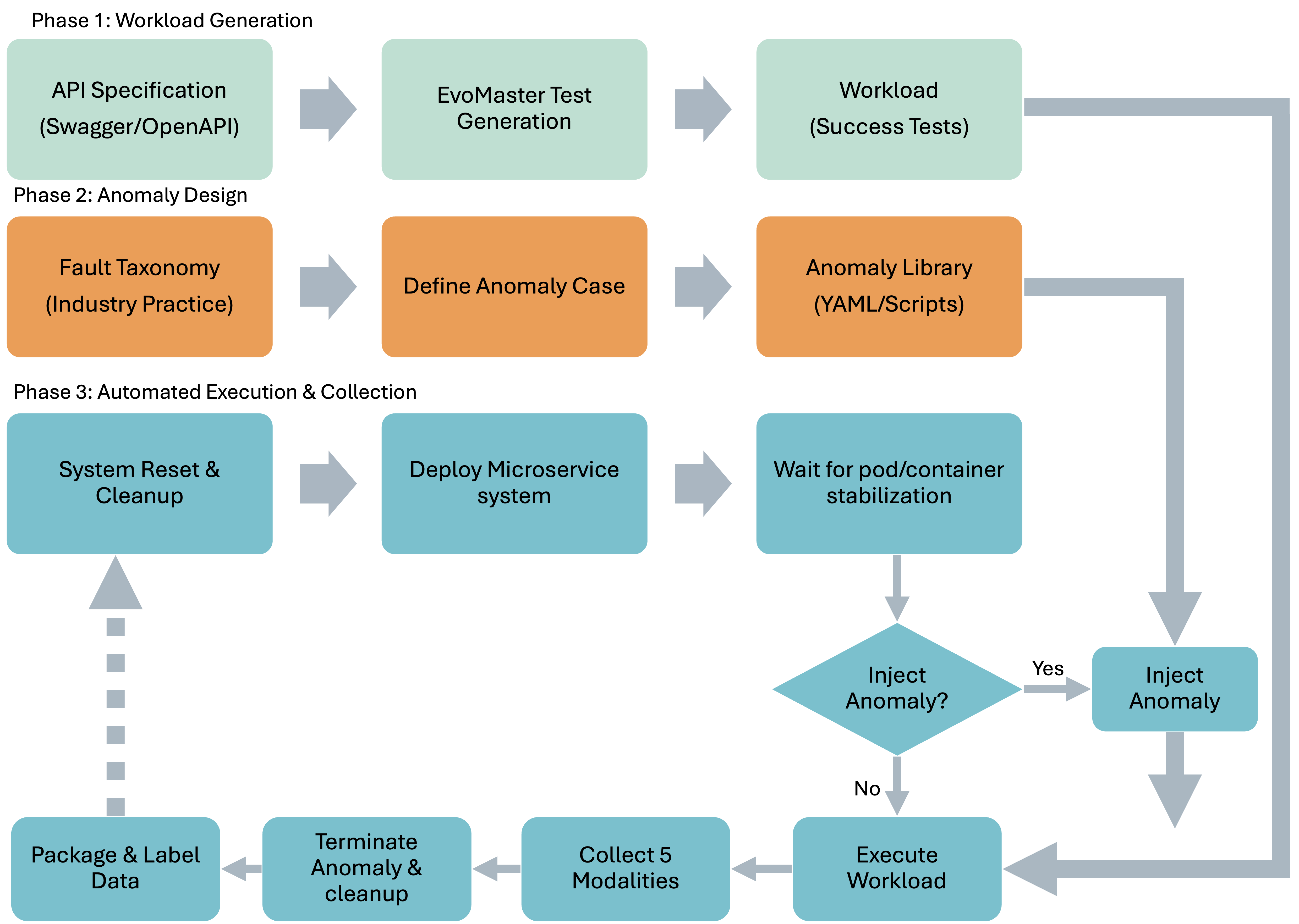}
  \caption{The three-phase automated data collection pipeline.}
  \label{fig:pipeline}
\end{figure}

Our data collection pipeline contains three phases, as illustrated in Figure~\ref{fig:pipeline}. It systematically generates workloads, injects anomalies, and captures multimodal data. We describe each phase in detail below. 

\textbf{Phase 1.} We create system workloads that invoke service endpoints to simulate realistic user requests through automated black-box testing using EvoMaster. EvoMaster \cite{andrea_arcuri_2025_16808704} is a popular open-source tool for automated black-box testing of RESTful and web services. It generates automated test suites based on API specifications (Swagger 2.0/OpenAPI 3.0) and uses evolutionary search and fuzzing algorithms to maximize API endpoint coverage. To support EvoMaster’s black-box testing, we generate API specifications for the target MSS. These specifications are parsed by EvoMaster to produce automated test suites. For data collection, we only utilize the “success tests” produced by EvoMaster. The fault tests and EvoMaster generated reports are used for specification diagnostics, hardening, and evolution. Anomalies are injected in the next phase of the collection pipeline, independently under controlled experimental conditions. 

 \textbf{Phase 2.} This phase prepares for and manages the controlled injection of anomalies, based on our four-level anomaly taxonomy (Table~\ref{tab:anomalies}). We predefine each anomaly case as a reusable configuration file or script, forming an Anomaly Library.
 We injected Performance, Service, Database level anomalies using ChaosMesh \cite{chaosmesh} via Kubernetes Custom Resource Definitions (CRDs) to orchestrate pod failures, network latency, and packet loss via the Linux kernel's traffic control (tc) subsystem. For Code-level anomalies, we leverages JVM sandbox technology from ChaosBlade \cite{chaosblade} to perform dynamic bytecode instrumentation at runtime. This allows to inject exceptions and delays into specific Java methods without modifying the source code or redeploying different binaries, reproducing logic defects in production artifacts.
These baseline parameters are widely used in industrial anomaly injection testing and ensure a balance between observability of system degradation and safe recovery after each experiment. Before executing the workload in Phase 3, the pipeline selects the appropriate anomaly case from this library based on the experiment plan. Importantly, the pipeline logic ensures the anomaly case is activated precisely before the workload starts and is terminated immediately upon workload completion. This strict timing provides clear temporal boundaries, isolating the anomaly's effects within the collected data for accurate analysis.

 \textbf{Phase 3.} We create a master script to ensure that each anomaly case run is isolated and starts from a clean system state via system reset and cleanup. To achieve this, the script redeploys a fresh instance of the MSS for each anomaly case run and waits until all components are fully initialized and stable. Since the original system logs only contained startup events, we inserted info-level normal logs into each microservice of SocialNetwork, with the aim of tracking the request processing flow and key operation results of each service during operation. Based on the experimental plan (input parameter), the script decides whether to inject an anomaly from the library (Phase 2). Subsequently, it executes the pre-generated workload (Phase 1) while concurrently capturing all five multimodal data through specific collectors. Upon workload completion, any active anomaly is terminated, the data collected for each case is saved in a folder named after the anomaly.

\section{Data Description}
The dataset primarily supports AIOps work on MSS, enabling researchers and practitioners to develop and evaluate AD \& RCA analysis techniques. The main contributions are the five correlated data modalities, including novel code coverage and API Res data, enabling the study of multimodal fusion and fine-grained analyses that track symptoms back to code execution and forward to user impact. 
The four level anomaly taxonomy also supports targeted evaluation across diverse, realistic anomaly types rather than only performance issues. Table~\ref{tab:anomalies} provides a detailed overview of the 24 anomaly cases, categorized by level, that we designed and injected across the SocialNetwork and TrainTicket systems.

\begin{table*}[t]
\centering
\caption{Taxonomy of Injected Anomaly Cases.}
\label{tab:anomalies}

\setlength{\tabcolsep}{4pt}
\renewcommand{\arraystretch}{1.05}

\begin{tabular}{@{}
  >{\raggedright\arraybackslash}p{1.9cm}   
  >{\raggedright\arraybackslash}p{5.3cm}   
  >{\raggedright\arraybackslash}p{10.0cm}  
@{}}
\toprule
\textbf{Level} & \textbf{Anomaly Cases}  & \textbf{Description} \\
\midrule
\multirow{3}{*}{\textbf{Performance}} 
& {CPU Contention}
&  High processor load on services or system wide. \\
&  {Network Performance Degradation} 
& Unstable network conditions such as packet loss or high latency. \\
&  {Disk I/O Stress}
&  Storage bottlenecks affecting data read/write operations. \\
\midrule
\multirow{3}{*}{\textbf{Service}} 
&  {Service Instance Failure}
&  Unexpected service crashes, terminations. \\
&  {Service Communication Failure}
& Services cannot reach their dependencies due to failed service discovery.\\
&  {Service Response Failure}
&  Service responding incorrectly. \\
\midrule
\multirow{3}{*}{\textbf{Database}} 
&  {Cache Pressure}
&  Memory pressure specifically targeting database caches. \\
&  {Connection Pool Exhaustion}
&  All database connections are occupied (via injected network delay). \\
&  {Transaction Issues}
&  Problems within database operations. \\
\midrule
\multirow{3}{*}{\textbf{Code}} 
&  {Manipulating Method Return Values}
&  Forcing specific Java methods in the application to return predefined values. \\
&  {Method Execution Delay}
&  Adding artificial delays to specific Java method calls. \\
&  {Injecting Exceptions}
&  Throwing specified exceptions during the execution of targeted Java methods. \\
\bottomrule
\end{tabular}%

\end{table*}

For SocialNetwork, we selected 5 services with 6 API paths for their deterministic, stateless benchmarking behavior and structured JSON responses. We excluded endpoints due to interactive workflows, session state, non-JSON media, or side effects. Dynamic analysis confirmed that automated clients target these selected endpoints. For TrainTicket, we targeted 34 services with 172 paths, excluding those with incomplete specifications, missing routes, or infrastructure roles.
\begin{table*}[bp!]
\centering
\caption{Scale and Composition of the Dataset}
\label{tab:dataset-description}
\begin{tabular}{@{}lcccccr@{}}
\toprule
\textbf{System} & \textbf{Log Lines (K)} & \textbf{Log Templates} & \textbf{Traces} & \textbf{Metrics Collected} & \textbf{API Requests} & \textbf{Average CCR} \\
\midrule
\textbf{TrainTicket} & 444.6 & 831 & 63,975 & 33 & 98,073 & 0.417 \\
\textbf{SocialNetwork} & 3958.5 & 29 & 2,635 & 29 & 1,950 & 0.682 \\
\bottomrule
\end{tabular}%
\end{table*}

We summarize our dataset in Table~\ref{tab:dataset-description}, reporting its scale and composition across modalities. 
“Log Lines” count the total number of lines produced after workload execution.
“Log Templates” denote structural patterns mined from raw logs \cite{8029742}.
The “Traces” column shows the total number of collected traces.
“Metrics Collected” count the number of unique Prometheus metrics.
“API Requests” count total requests across all EvoMaster runs for the system.
“Average CCR” denotes the mean code coverage rate computed over all coverage reports.

\section{Conclusion}

We introduced a new multimodal benchmark dataset for AD \& RCA in MSS. We added two novel modalities: API Res, which provides more user-facing information, and CCR, which enables fine-grained root cause analysis for software developers.
We believe this dataset will benefit the development and evaluation of advanced AIOps techniques. The dataset is publicly available to foster further research in building more resilient MSS.

The limitations are that only two open-source systems were studied and that anomalies were injected solely using chaos engineering tools, which may not fully mimic the complexity of all production environments. Additionally, the workload is only generated by EvoMaster’s black-box testing. In the future, we may use different workload generation tools and white-box testing approaches. 

\section{Acknowledgment}
This work has been supported by FAST, the Finnish Software Engineering Doctoral Research Network, funded by the Ministry of Education and Culture, Finland. This work is also funded by the EuroHPC Joint Undertaking and its members including top-up funding by the Ministry of Education and Culture of Finland. The authors acknowledge CSC-IT Center for Science, Finland, for providing computing resources. This work is also supported by the Research Council of Finland (grant id: 359861, the MuFAno project).

\clearpage

\bibliographystyle{ACM-Reference-Format}
\bibliography{reference}

@String{Computing = "Computing" }

@String{Springer = "Springer-Verlag" }

@inproceedings{10.1145/3643991.3644890,
author = {Amoroso d'Aragona, Dario and Bakhtin, Alexander and Li, Xiaozhou and Su, Ruoyu and Adams, Lauren and Aponte, Ernesto and Boyle, Francis and Boyle, Patrick and Koerner, Rachel and Lee, Joseph and Tian, Fangchao and Wang, Yuqing and Nyyss\"{o}l\"{a}, Jesse and Quevedo, Ernesto and Rahaman, Shahidur Md and Abdelfattah, Amr S. and M\"{a}ntyl\"{a}, Mika and Cerny, Tomas and Taibi, Davide},
title = {A Dataset of Microservices-based Open-Source Projects},
year = {2024},
isbn = {9798400705878},
publisher = {Association for Computing Machinery},
address = {New York, NY, USA},
url = {https://doi.org/10.1145/3643991.3644890},
doi = {10.1145/3643991.3644890},
booktitle = {Proceedings of the 21st International Conference on Mining Software Repositories},
pages = {504–509},
numpages = {6},
location = {Lisbon, Portugal},
series = {MSR '24}
}

@article{10.1145/3715742,
author = {Wang, Yuqing and M\"{a}ntyl\"{a}, Mika V. and Demeyer, Serge and Beyaz\i{}t, Mutlu and Kisaakye, Joanna and Nyyss\"{o}l\"{a}, Jesse},
title = {Cross-System Categorization of Abnormal Traces in Microservice-Based Systems via Meta-Learning},
year = {2025},
issue_date = {July 2025},
publisher = {Association for Computing Machinery},
address = {New York, NY, USA},
volume = {2},
number = {FSE},
url = {https://doi.org/10.1145/3715742},
doi = {10.1145/3715742},
journal = {Proc. ACM Softw. Eng.},
month = jun,
articleno = {FSE027},
numpages = {23}
}

@INPROCEEDINGS{10992557,
  author={Wang, Yuqing and Mäntylä, Mika V. and Nyyssölä, Jesse and Ping, Ke and Wang, Liqiang},
  booktitle={2025 IEEE International Conference on Software Analysis, Evolution and Reengineering (SANER)}, 
  title={Cross-System Software Log-based Anomaly Detection Using Meta-Learning}, 
  year={2025},
  volume={},
  number={},
  pages={454-464},
  doi={10.1109/SANER64311.2025.00049}}

@INPROCEEDINGS{10803340,
  author={Steidl, Monika and Dornauer, Benedikt and Felderer, Michael and Ramler, Rudolf and Racasan, Mircea-Cristian and Gattringer, Marko},
  booktitle={2024 50th Euromicro Conference on Software Engineering and Advanced Applications (SEAA)}, 
  title={How Industry Tackles Anomalies during Runtime: Approaches and Key Monitoring Parameters}, 
  year={2024},
  volume={},
  number={},
  pages={364-372},
  doi={10.1109/SEAA64295.2024.00062}}

@ARTICLE{8580420,
  author={Zhou, Xiang and Peng, Xin and Xie, Tao and Sun, Jun and Ji, Chao and Li, Wenhai and Ding, Dan},
  journal={IEEE Transactions on Software Engineering}, 
  title={Fault Analysis and Debugging of Microservice Systems: Industrial Survey, Benchmark System, and Empirical Study}, 
  year={2021},
  volume={47},
  number={2},
  pages={243-260},
  doi={10.1109/TSE.2018.2887384}}

@article{10.1145/3501297,
author = {Soldani, Jacopo and Brogi, Antonio},
title = {Anomaly Detection and Failure Root Cause Analysis in (Micro) Service-Based Cloud Applications: A Survey},
year = {2022},
issue_date = {March 2023},
publisher = {Association for Computing Machinery},
address = {New York, NY, USA},
volume = {55},
number = {3},
issn = {0360-0300},
url = {https://doi.org/10.1145/3501297},
doi = {10.1145/3501297},
journal = {ACM Comput. Surv.},
month = feb,
articleno = {59},
numpages = {39}
}

@article{li2022constructing,
  title={Constructing large-scale real-world benchmark datasets for aiops},
  author={Li, Zeyan and Zhao, Nengwen and Zhang, Shenglin and Sun, Yongqian and Chen, Pengfei and Wen, Xidao and Ma, Minghua and Pei, Dan},
  journal={arXiv preprint arXiv:2208.03938},
  year={2022}
}

@inproceedings{10.1145/3701716.3715290,
author = {Pham, Luan and Zhang, Hongyu and Ha, Huong and Salim, Flora and Zhang, Xiuzhen},
title = {RCAEval: A Benchmark for Root Cause Analysis of Microservice Systems with Telemetry Data},
year = {2025},
isbn = {9798400713316},
publisher = {Association for Computing Machinery},
address = {New York, NY, USA},
url = {https://doi.org/10.1145/3701716.3715290},
doi = {10.1145/3701716.3715290},
booktitle = {Companion Proceedings of the ACM on Web Conference 2025},
pages = {777–780},
numpages = {4},
location = {Sydney NSW, Australia},
series = {WWW '25}
}

@inproceedings{10.1145/3297858.3304013,
author = {Gan, Yu and Zhang, Yanqi and Cheng, Dailun and Shetty, Ankitha and Rathi, Priyal and Katarki, Nayan and Bruno, Ariana and Hu, Justin and Ritchken, Brian and Jackson, Brendon and Hu, Kelvin and Pancholi, Meghna and He, Yuan and Clancy, Brett and Colen, Chris and Wen, Fukang and Leung, Catherine and Wang, Siyuan and Zaruvinsky, Leon and Espinosa, Mateo and Lin, Rick and Liu, Zhongling and Padilla, Jake and Delimitrou, Christina},
title = {An Open-Source Benchmark Suite for Microservices and Their Hardware-Software Implications for Cloud \& Edge Systems},
year = {2019},
isbn = {9781450362405},
publisher = {Association for Computing Machinery},
address = {New York, NY, USA},
url = {https://doi.org/10.1145/3297858.3304013},
doi = {10.1145/3297858.3304013},
booktitle = {Proceedings of the Twenty-Fourth International Conference on Architectural Support for Programming Languages and Operating Systems},
pages = {3–18},
numpages = {16},
location = {Providence, RI, USA},
series = {ASPLOS '19}
}

@INPROCEEDINGS{10172617,
  author={Lee, Cheryl and Yang, Tianyi and Chen, Zhuangbin and Su, Yuxin and Lyu, Michael R.},
  booktitle={2023 IEEE/ACM 45th International Conference on Software Engineering (ICSE)}, 
  title={Eadro: An End-to-End Troubleshooting Framework for Microservices on Multi-source Data}, 
  year={2023},
  volume={},
  number={},
  pages={1750-1762},
  doi={10.1109/ICSE48619.2023.00150}}

@inproceedings{10.1145/3611643.3616249,
author = {Yu, Guangba and Chen, Pengfei and Li, Yufeng and Chen, Hongyang and Li, Xiaoyun and Zheng, Zibin},
title = {Nezha: Interpretable Fine-Grained Root Causes Analysis for Microservices on Multi-modal Observability Data},
year = {2023},
isbn = {9798400703270},
publisher = {Association for Computing Machinery},
address = {New York, NY, USA},
url = {https://doi.org/10.1145/3611643.3616249},
doi = {10.1145/3611643.3616249},
booktitle = {Proceedings of the 31st ACM Joint European Software Engineering Conference and Symposium on the Foundations of Software Engineering},
pages = {553–565},
numpages = {13},
location = {San Francisco, CA, USA},
series = {ESEC/FSE 2023}
}

@inproceedings{10.1145/3727582.3728682,
author = {Bakhtin, Alexander and Nyyss\"{o}l\"{a}, Jesse and Wang, Yuqing and Ahmad, Noman and Ping, Ke and Esposito, Matteo and M\"{a}ntyl\"{a}, Mika and Taibi, Davide},
title = {LO2: Microservice API Anomaly Dataset of Logs and Metrics},
year = {2025},
isbn = {9798400715945},
publisher = {Association for Computing Machinery},
address = {New York, NY, USA},
url = {https://doi.org/10.1145/3727582.3728682},
doi = {10.1145/3727582.3728682},
booktitle = {Proceedings of the 21st International Conference on Predictive Models and Data Analytics in Software Engineering},
pages = {1–10},
numpages = {10},
location = {Trondheim, Norway},
series = {PROMISE '25}
}

@InProceedings{10.1007/978-3-030-44769-4_13,
author="Nedelkoski, Sasho
and Bogatinovski, Jasmin
and Mandapati, Ajay Kumar
and Becker, Soeren
and Cardoso, Jorge
and Kao, Odej",
editor="Brogi, Antonio
and Zimmermann, Wolf
and Kritikos, Kyriakos",
title="Multi-source Distributed System Data for AI-Powered Analytics",
booktitle="Service-Oriented and Cloud Computing",
year="2020",
publisher="Springer International Publishing",
address="Cham",
pages="161--176",
isbn="978-3-030-44769-4"
}

@INPROCEEDINGS{8029742,
  author={He, Pinjia and Zhu, Jieming and Zheng, Zibin and Lyu, Michael R.},
  booktitle={2017 IEEE International Conference on Web Services (ICWS)}, 
  title={Drain: An Online Log Parsing Approach with Fixed Depth Tree}, 
  year={2017},
  volume={},
  number={},
  pages={33-40},
  doi={10.1109/ICWS.2017.13}}

@article{10.1145/3715005,
author = {Zhang, Shenglin and Xia, Sibo and Fan, Wenzhao and Shi, Binpeng and Xiong, Xiao and Zhong, Zhenyu and Ma, Minghua and Sun, Yongqian and Pei, Dan},
title = {Failure Diagnosis in Microservice Systems: A Comprehensive Survey and Analysis},
year = {2025},
publisher = {Association for Computing Machinery},
address = {New York, NY, USA},
issn = {1049-331X},
url = {https://doi.org/10.1145/3715005},
doi = {10.1145/3715005},
note = {Just Accepted},
journal = {ACM Trans. Softw. Eng. Methodol.},
month = jan
}

@misc{void-2024-report,
  author       = {{Verica, Inc.}},
  title        = {The Verica Open Incident Database (VOID) Report 2024},
  year         = {2024},
  doi          = {10.5281/zenodo.13499121},
  url          = {https://www.thevoid.community/report-2024},
  note         = {Accessed: 2025-10-15}
}

@misc{aws-well-architected-reliability-2024,
  author = {{Amazon Web Services}},
  title  = {Reliability Pillar - AWS Well-Architected Framework},
  year   = {2024},
  url    = {https://docs.aws.amazon.com/pdfs/wellarchitected/latest/reliability-pillar/wellarchitected-reliability-pillar.pdf},
  note   = {Publication date: 2024-11-06; Accessed: 2025-10-15}
}

@misc{azure-failure-mode-analysis,
  author = {{Microsoft Azure}},
  title  = {Architecture strategies for performing failure mode analysis},
  year   = {2024},
  url    = {https://learn.microsoft.com/en-us/azure/well-architected/reliability/failure-mode-analysis},
  note   = {Accessed: 2025-10-15}
}

@misc{azure-service-fabric-testability,
  author = {{Microsoft Azure}},
  title  = {Simulate failures in Azure microservices (Service Fabric Testability)},
  year   = {2024},
  url    = {https://learn.microsoft.com/en-us/azure/service-fabric/service-fabric-testability-actions},
  note   = {Accessed: 2025-10-15}
}

@misc{prometheus,
  author       = {{Prometheus Authors}},
  title        = {Prometheus: Monitoring system \& time series database},
  year         = {2025},
  url          = {https://prometheus.io/},
  lastaccessed = {October 20, 2025}
}

@misc{jaeger,
  author       = {{Jaeger Authors}},
  title        = {Jaeger: Open Source Distributed Tracing Platform},
  year         = {2025},
  url          = {https://www.jaegertracing.io/},
  lastaccessed = {2025-10-20}
}

@misc{skywalking,
  author       = {{Apache SkyWalking}},
  title        = {Apache SkyWalking: APM and Observability Platform},
  year         = {2025},
  url          = {https://skywalking.apache.org/},
  lastaccessed = {2025-10-20}
}

@misc{gnu-gcov-2024,
  author       = {{Free Software Foundation}},
  title        = {Gcov---a Test Coverage Program},
  year         = {2024},
  url          = {https://gcc.gnu.org/onlinedocs/gcc/Gcov.html},
  note         = {Accessed: 2025-10-20}  
}

@misc{jacoco,
  author       = {{The JaCoCo Team}},
  title        = {JaCoCo -- Java Code Coverage Library},
  year         = {2025},
  url          = {https://www.jacoco.org/},
  note         = {Accessed: 2025-10-20}
}

@software{andrea_arcuri_2025_16808704,
  author       = {Andrea Arcuri and
                  Man Zhang and
                  Susruthan Seran and
                  Asma Belhadi and
                  Juan Pablo Galeotti and
                  Bogdan and
                  Amid Golmohammadi and
                  Onur Duman and
                  Agustina Aldasoro and
                  Philip and
                  Ömür Şahin and
                  Alberto Martín López and
                  Hernan Ghianni and
                  Mohsen Taheri Shalmani and
                  Miguel Rodriguez and
                  IvaK and
                  Annibale Panichella and
                  Kyle Niemeyer and
                  Marcello Maugeri},
  title        = {WebFuzzing/EvoMaster: v4.0.0},
  month        = aug,
  year         = 2025,
  publisher    = {Zenodo},
  version      = {v4.0.0},
  doi          = {10.5281/zenodo.16808704},
  url          = {https://doi.org/10.5281/zenodo.16808704},
}

@inproceedings{10.1145/3510003.3510180,
author = {Zhang, Chenxi and Peng, Xin and Sha, Chaofeng and Zhang, Ke and Fu, Zhenqing and Wu, Xiya and Lin, Qingwei and Zhang, Dongmei},
title = {DeepTraLog: trace-log combined microservice anomaly detection through graph-based deep learning},
year = {2022},
isbn = {9781450392211},
publisher = {Association for Computing Machinery},
address = {New York, NY, USA},
url = {https://doi.org/10.1145/3510003.3510180},
doi = {10.1145/3510003.3510180},
booktitle = {Proceedings of the 44th International Conference on Software Engineering},
pages = {623–634},
numpages = {12},
location = {Pittsburgh, Pennsylvania},
series = {ICSE '22}
}

@misc{chaosmesh,
  author = {{Chaos Mesh Authors}},
  title  = {Chaos Mesh: A Chaos Engineering Platform for Kubernetes},
  year   = {2025},
  url    = {https://chaos-mesh.org/},
  lastaccessed = {October 16, 2025}
}

@misc{chaosblade,
  author       = {{ChaosBlade Authors}},
  title        = {ChaosBlade: Cloud-Native Chaos Engineering Toolkit},
  year         = {2025},
  url          = {https://chaosblade.io/en/},
  lastaccessed = {October 16, 2025}
}

\end{document}